\documentclass[aps,groupedaddres8s,fleqn,nofootinbib,twocolumn]{revtex4}
\usepackage{graphicx}
\usepackage{xcolor}
\usepackage{amsmath,amssymb}
\usepackage{float}
\usepackage{amsfonts}
\usepackage{verbatim}
\usepackage{flushend}
\usepackage{balance}
\usepackage{endnotes}
\usepackage{footnote}
\usepackage{adjustbox}
\usepackage{gensymb}
\usepackage{float}
\usepackage{epigraph}
\usepackage{xcolor}
\usepackage[utf8]{inputenc}
\usepackage{setspace}

\newcommand{\beq}{\begin{eqnarray}}
\newcommand{\eeq}{\end{eqnarray}}
\usepackage{amsmath}

\begin{document}

\title{Microscopic dynamics and Bose-Einstein condensation in liquid helium}
\author{K. Trachenko$^{1}$}
\address{$^1$ School of Physical and Chemical Sciences, Queen Mary University of London, Mile End Road, London, E1 4NS, UK}

\begin{abstract}
We review fundamental problems involved in liquid theory including both classical and quantum liquids. Understanding classical liquids involves exploring details of their microscopic dynamics and
its consequences. Here, we apply the same general idea to quantum liquids. We discuss momentum condensation in liquid helium which is consistent with microscopic dynamics in liquids and high
mobility of liquid atoms. We propose that mobile transit atoms accumulate in the finite-energy state where the transit speed is close to the speed of sound. In this state, the transit energy is
close to the zero-point energy. In momentum space, the accumulation operates on a sphere with the radius set by interatomic spacing and corresponds to zero net momentum. We show that this picture
is supported by experiments, including the kinetic energy of helium atoms below the superfluid transition and sharp peaks of scattered intensity at predicted energy. We discuss the implications of
this picture including the macroscopic wave function and superfluidity. The superfluid transition temperature is evaluated to be within 15\% of the experimental value.
\end{abstract}

\maketitle

\section{Introduction}

Problems involved in liquid theory are fundamental. As discussed by Landau, Lifshitz and Pitaevskii (LLP), these problems include (a) strong interatomic interactions combined with dynamical disorder and (b) the absence of a small parameter \cite{landaustat,pita1}. For this reason, no general theory of liquids was long thought to be possible, in contrast to theories of solids and gases. For example, calculating generally-applicable thermodynamic properties such as energy and heat capacity as well as their temperature dependence  form essential part of theories of solids and gases. Deriving such general relations was considered impossible in liquids \cite{landaustat,pita1}.

The first part of the problem stated by LLP can be illustrated by writing the liquid energy as

\begin{equation}
E=\frac{3}{2}NT+\frac{n}{2}\int g(r)u(r)dV
\label{enint}
\end{equation}

\noindent where $n$ is concentration, $g(r)$ is the pair distribution function, $u(r)$ is the interaction potential, interactions and correlations are assumed to be pairwise and $k_{\rm B}=1$.

Early liquid theories  \cite{kirkwoodbook,borngreen,zwanzig,barkerhenderson} considered that the goal of the statistical theory of liquids is to provide a relation between liquid thermodynamics and liquid structure and intermolecular interactions such as $g(r)$ and $u(r)$ in Eq. \eqref{enint}. Working towards this goal involved developing the analytical models for liquid structure and interactions, which has become the essence of these theories \cite{kirkwoodbook,borngreen,zwanzig,barkerhenderson,egelstaff,faber,march,marchtosi,tabor,balucani,hansen2,hansen1}. The problem is that the interaction $u(r)$ in liquids is both strong and system-specific, hence $E$ in Eq. \eqref{enint} is strongly system-dependent as stated by LLP. For this reason, no generally applicable theory of liquids was considered possible \cite{landaustat,pita1}. An additional difficulty is that interatomic interactions and correlation functions are generally not available apart from fairly simple model liquids and can be generally complex involving many-body, hydrogen-bond interactions and so on. This precludes calculation of the liquid energy in the approach based on Eq. \eqref{enint} or its extensions involving, for example, higher-order correlation functions \cite{borngreen,barkerhenderson}. Even when $g(r)$ and $u(r)$ are available in simple cases, the calculation involving Eq. \eqref{enint} or similar is not enough: in order to explain experimental temperature dependence of energy and heat capacity of real liquids \cite{ropp,wallacecv,wallacebook,proctor1,proctor2,chen-review}, one still needs to develop a physical model in this approach.

Common liquid models are inapplicable to understanding the energy and heat capacity of real liquids. These models include the widely discussed Van der Waals model, the hard-spheres model and their extensions \cite{hansen2,ziman,march,parisihard}. Both models give the specific heat $c_v=\frac{3}{2}k_{\rm B}$ \cite{landaustat,wallacecv,wallacebook}, the ideal-gas value, in contrast to experiments showing liquid $c_v=3k_{\rm B}$ close to melting \cite{wallacecv,wallacebook,proctor2}. These models were also used as reference states to calculate the energy \eqref{enint} by expanding interactions into repulsive and attractive parts (see, e.g., Refs. \cite{barkerhenderson,wca1,wca2,zwanzig,rosentar}). These parts understandably play different roles at high and low density, however this method faces the same problem outlined by LLP: the interactions and expansion coefficients are strongly system-dependent and so are the final results, precluding a general theory.

In solids, both crystalline and amorphous, the above issues do not emerge because the solid state theory is based on collective excitations, phonons. This theory is predictive, physically transparent and generally applicable to all solids. There is no need to explicitly consider structure and interactions in order to understand basic thermodynamic properties of solids. Most important results such as universal temperature dependence of energy and heat capacity readily come out in the phonon approach to solids \cite{landaustat}.

The simplifying small parameter in solids is small phonon displacements from equilibrium, but this seemingly does not apply to liquids because liquids do not have stable equilibrium points that can be used to sustain these small phonon displacements. Weakness of interactions used in the theory of gases does not apply to liquids either because interactions in liquids are as strong as in solids. This constitutes the second, no small parameter problem, outlined by LLP.

Working towards solving the liquid theory problem involved several steps. The important first step involved the consideration of microscopic dynamics of liquid particles provided by the Frenkel theory \cite{frenkel}: differently from solids where particle dynamics is purely oscillatory and gases where dynamics is purely diffusive, particle dynamics in liquids is mixed and combines oscillations around quasi-equilibrium points as in solids and diffusive motions between different points. This consideration led to understanding collective excitations in liquids, phonons, with an important property that the phase space available to these phonons is not fixed as in solids but is instead variable \cite{ropp,proctor1,proctor2,chen-review}. In particular, this phase space reduces with temperature. This effect has a general implication: specific heat of classical liquids universally decreases with temperature, in agreement with experiments \cite{ropp,proctor1,proctor2} (in molecular liquids, a competing increase of $c_v$ at high temperature may be related to progressive excitations of internal molecular degrees of freedom - this mechanism is also present in quantum solids and is unrelated to liquid physics). This effect also addresses the no small parameter problem stated by Landau, Lifshitz and Pitaevskii: the small parameter exists in liquids but it operates in a variable phase space.

The above theory is closely based on details of microscopic particle dynamics in liquids. This was essential to understand classical liquids. In this paper, we explore the implications of microscopic dynamics in quantum liquids. There is a good reason for doing this: a microscopic theory requires microscopic dynamics. For example, understanding superconductivity is aided by its microscopic theory \cite{bcs}. In contrast, no microscopic theory of liquid helium $^4$He and its superfluid properties is considered to exist, as noted by several observers. For example, Pines and Nozieres observe \cite{pinesnoz} that ``microscopic theory does not at present provide a quantitative description of liquid He II'' (``II'' here refers to helium below the superfluid transition temperature $T_\lambda=2.17$ K), adding that a microscopic theory exists only for models of dilute gases or models with weak interactions where perturbation theory applies such as the Bogoliubov theory. Griffin similarly remarks that we can't make quantitative predictions of superfluid $^4$He on the basis of existing theories and depend on experimental data for guidance, and attributes theoretical problems to the ``difficulties of dealing with a liquid, whether Bose-condensed or not'' \cite{griffin}. Pines, Nozieres, Griffin and other authors recognize that the same fundamental theoretical problems outlined by LLP and related to strong interactions combined with dynamical disorder apply to both quantum and classical liquids.

In view of this, progress in the field is likely to come from discussing microscopic details of liquid helium and particle dynamics in particular. This dynamics is a starting point of our discussion.

In liquid helium, atoms are located in the range of the strong part of the potential and close to its minimum as follows from, for example, pair distribution function \cite{ceperley}. Since the early
work of Frenkel, we know the key element of microscopic dynamics in such a liquid at low temperature: each particle oscillates around quasi-equilibrium position in a ``cage'', followed by the jump to
the neighbouring position \cite{frenkel}. Path integral simulations show minima of the velocity autocorrelation function below $T_\lambda$ in liquid helium \cite{path-he}, indicating the presence of
the oscillatory component of particle motion \cite{flreview}. The oscillations are interrupted by atomic jumps separated in time by liquid relaxation time $\tau$ \cite{frenkel}. Seen in molecular
dynamics simulations and referred to as ``transits'' \cite{wallacemd}, these mobile atoms give liquids their ability to flow. This applies to both classical and quantum liquids.

This oscillatory-transit dynamics consistently explains dynamical and thermodynamic properties of liquids \cite{frenkel,dyre,ropp,proctor2}. Once considered in liquid helium, this dynamics presents a problem for understanding Bose-Einstein condensation (BEC), as follows.

Understanding liquid  $^4$He involved exciting new ideas including attributing superfluidity to the BEC in zero momentum state $p=0$ where only about 10\% of atoms are in the condensate
\cite{pinesnoz,griffin,ceperley,pitaevs,annett,leggett,glyde,balibar}. The experimental evidence for this is not unambiguous (see, for example, Ref. \cite{temp-energy4}, p. 175 and cited references
therein as well as discussion in Sections \ref{exp} and \ref{psi}), however there is also an issue related to microscopic dynamics. In gases, BEC understandably corresponds to the immobile state
$p=0$ at low temperature. Liquids are importantly different because every atom in the liquid necessarily possesses high mobility at any temperature: an atom either oscillates with the ever-present
component of zero-point motion (this motion is particularly important in liquid $^4$He \cite{annett}) or becomes a transit \cite{frenkel,dyre,wallacemd,proctor2}. As in solids, the oscillatory atoms
do not undergo the BEC. This leaves the flow-enabling transits as the remaining sub-system where BEC can possibly operate. However, transits are inherently mobile and, moreover, strongly and
constantly interact with surrounding mobile atoms. Depending on the dynamical regime, these surrounding mobile atoms are either (a) both oscillatory and other mobile transit atoms in the
low-temperature oscillatory-transit regime of liquid dynamics or (b) mobile transit atoms in the high-temperature diffusive regime where the oscillatory component of motion is lost and all atoms
become mobile transits \cite{flreview,proctor2}. An atom in the condensate at $p=0$ can not remain immobile while strongly and constantly interacting with surrounding mobile atoms. Assuming 10\% of
atoms in the immobile $p=0$ state would imply that each {\it immobile} atom interacts with about 10 of its nearest-neighbour {\it mobile} atoms on average, with the interaction strength comparable to
that in solids. This is not possible. One could try to imagine that 10\% of atoms in the state $p=0$ are organised in clusters and hence are not affected by interactions with mobile atoms. However
this would apply to the bulk of the cluster only, whereas atoms at the surface of the cluster would still be excited by interactions with surrounding mobile atoms.

A somewhat related problem with BEC at $p=0$ in liquid helium was noted in the early studies of superfluidity by Landau \cite{landaumigdal}: nothing prevents atoms in BEC at $p=0$ to collide with excited atoms, leading to friction and hence no superfluidity. This problem was later described in terms of condensate depletion due to interactions \cite{griffin}, however it was not demonstrated that this depletion cannot be complete (see, e.g., p. 175 of Ref. \cite{temp-energy4}) as discussed in more detail below.

This concludes our longer-than-usual introduction. This was needed to put liquid helium in the wider context of a more general liquid research and liquid problem. This discussion leads us to look for a way to discuss liquid helium which is consistent with microscopic liquid dynamics and high atomic mobility.

In this paper, we discuss BEC in liquid helium which is consistent with microscopic dynamics in liquids and high mobility of liquid atoms in particular. We propose that a large number of transits
accumulate in the finite-energy state where the transit speed is close to the speed of sound.  In this state, the transit energy is close to the zero-point (vacuum) energy. In momentum space, the
accumulation operates on a {\it Bose-liquid sphere} with the radius set by interatomic spacing and corresponds to zero net momentum. We show that this picture is supported by experiments, including
the measured kinetic energy of helium atoms below the superfluid transition and sharp peaks  of scattered intensity at predicted energy. We discuss the implications of this picture including the
macroscopic wave function and superfluidity. We evaluate the superfluid transition temperature to be within 15\% of the experimental value.

\section{Results and discussion}

\subsection{Transit dynamics}
\label{dynamics}

Mobile transit atoms are relevant to superfluidity because they enable liquid flow and set its properties. At fixed cage volume, a transit involves surmounting large energy barrier due to steep interatomic repulsion and hence very long waiting times. Instead, the transit is enabled by temporary increase of the cage volume due to fluctuations, thermal or quantum, as proposed by Frenkel \cite{frenkel} and widely appreciated since \cite{dyre}. This is illustrated in Figure 1: when neighbouring atoms move out of the way, the transit to the neighbouring cage takes place where the oscillation resumes until the next transit, and so on.

The transit is a quantum-mechanical object: the de Broglie wavelength at 4 K where helium condenses is about 4 \AA, and is larger than the interatomic separation \cite{annett}. To write the
wavefunction of the transit, we consider its translational motion between two quasi-equilibrium positions in Figure 1. The transit is a fast process lasting during the shortest time scale in the
system close to the Debye vibration period, $\tau_{\rm D}$ \cite{frenkel,dyre}. The potential $V(x)$ acting on the transit from surrounding particles can be assumed as slowly-varying during the fast
transit process so that the Wentzel–Kramers–Brillouin approximation applies:

\begin{figure}
\begin{center}
{\scalebox{0.4}{\includegraphics{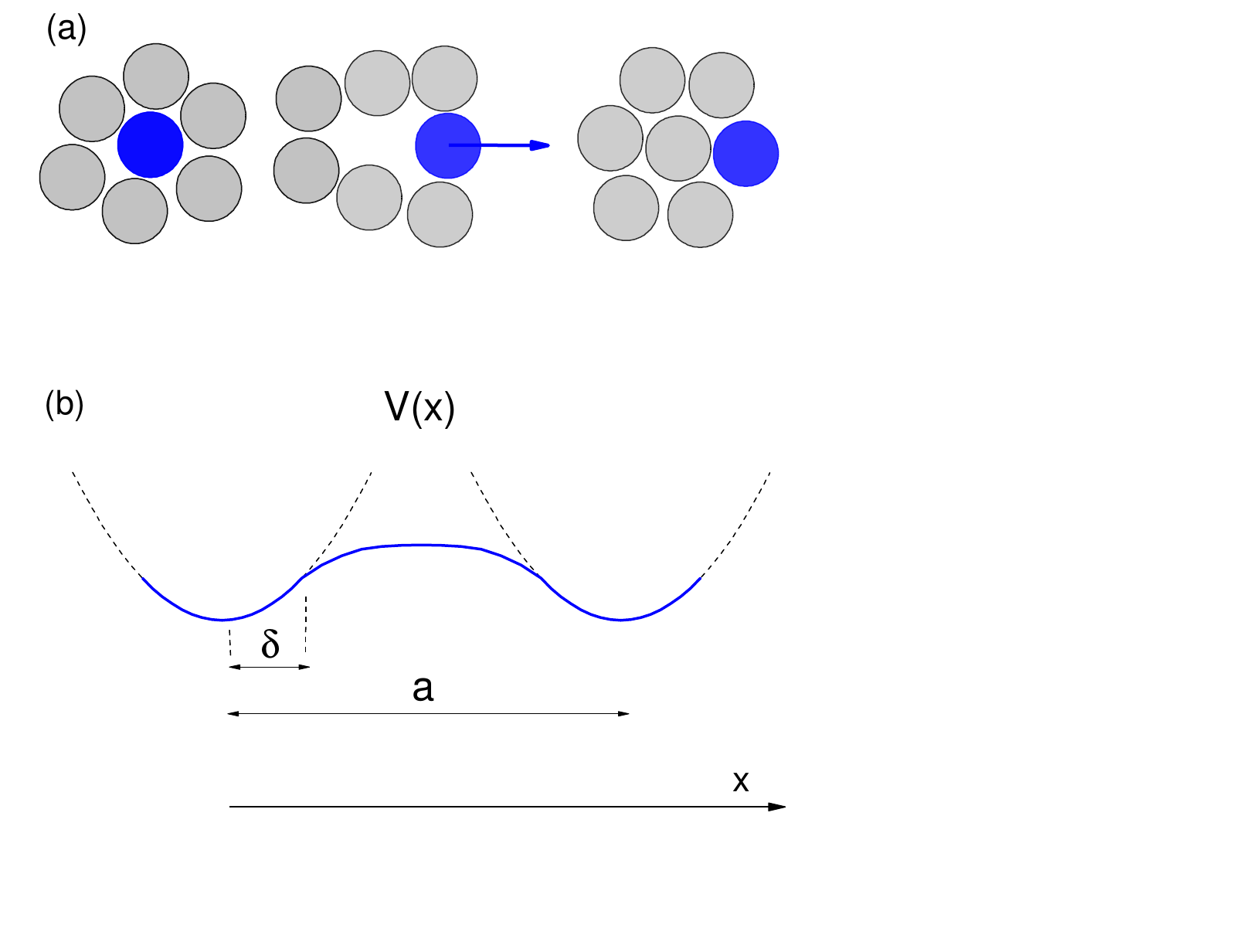}}}
{\scalebox{0.25}{\includegraphics{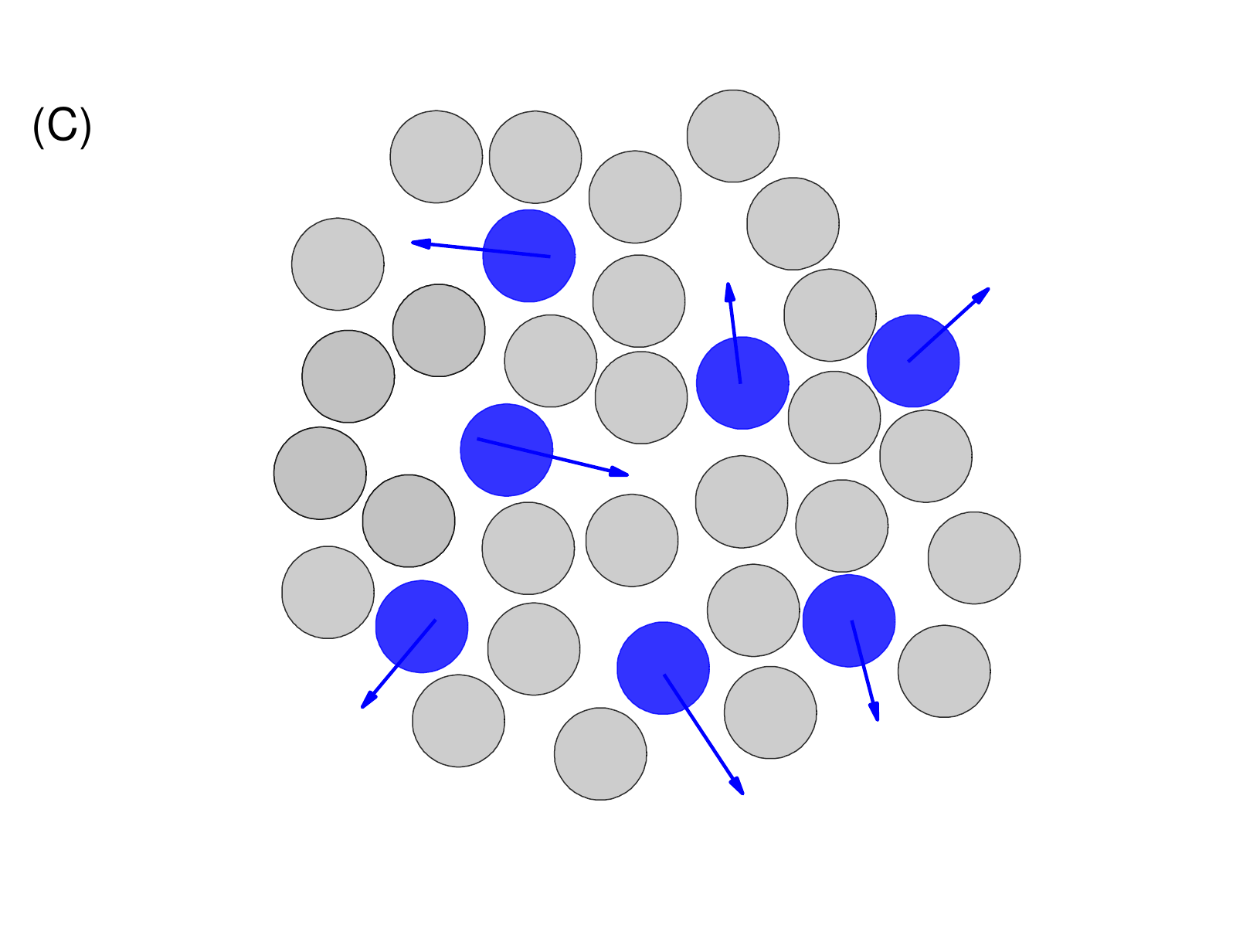}}}
\end{center}
\caption{(a) A mobile transit showing the diffusive transit motion between two quasi-equilibrium oscillatory states when the cage opens up due to fluctuations. (b) Potential experienced by the atom in the oscillatory and transit states. (c) A set of transit atoms in a liquid. Schematic illustration.}
\label{phase}
\end{figure}

\begin{eqnarray}
\begin{split}
& \psi(x)=\frac{1}{\sqrt{k(x)}}\exp\left(i\int\limits^x k(x)dx\right)\\
& k(x)=\left(\frac{2m}{\hbar^2}\left(E-V(x)\right)\right)^\frac{1}{2}
\end{split}
\label{wkb}
\end{eqnarray}

\noindent where $\psi(x)$ is the wavefunction of the transit and $E$ is the transit energy.

Approximating the slowly-varying potential experienced by the transit by a constant gives and assuming $E>V$ the plane wave with constant $k$: $\psi(x)\propto\exp\left(ikx\right)$. This wave function
continuously joins the oscillatory wavefunctions of the atom in the oscillatory states before and after the transit. An example of a quantum-mechanical model of the oscillatory-transit dynamics with
continuous solutions in space and time is shown in the Appendix.

A different possibility for the transit to move is to tunnel, the dominant mode of translational motion at low temperature in quantum liquids \cite{andreev}. In this case, $\tau$ is set by the
tunneling probability. We don't go into details of this difference here and consider the transit motion as a translational motion different from the oscillatory motion inside the cage.

We now make the key observation regarding the absolute value of transit momentum ${\bf p}_t$, $p_t=|{\bf p}_t|$. For a large number of atoms, $p_t$ is not arbitrary but is set by liquid structure and dynamics. The transit speed can be estimated as $v=\frac{\Delta x}{\Delta t}$, where $\Delta x$ is the distance travelled by the transit and $\Delta t$ is the transit time. $\Delta x$ is close to the interatomic separation in the liquid, $a$ (the ``UV'' cutoff in condensed matter). $\Delta t$ is set by atom inertia and the shortest time scale in the system comparable to the Debye vibration period $\tau_{\rm D}$. The mean value of $v$ (mean in a sense of $\Delta x$ fluctuating around $a$ and $\Delta t$ fluctuating around $\tau_{\rm D}$) is $v=\frac{a}{\tau_{\rm D}}$ or $v=a\omega_{\rm D}$, where $\omega_{\rm D}$ is the Debye frequency which is inverse of $\tau_{\rm D}$ and is close to the intra-cage rattling frequency. This ratio, $\frac{a}{\tau_{\rm D}}$, is the speed of sound, $c$. This follows from writing the dispersion relation in the linear form as $\omega=ck$, using $\omega=\omega_{\rm D}$ and Debye wavevector $k=k_{\rm D}$ and noting $\omega\approx\frac{1}{\tau_{\rm D}}$ and $k_{\rm D}\approx\frac{1}{a}$.

Hence the mean $p_t$ and energy $E_t$ of transits are close to

\begin{eqnarray}
\begin{split}
& p_t=mc\\
& E_t=\frac{mc^2}{2}=14~{\rm K}=0.28~{\rm THz}
\end{split}
\label{moment}
\end{eqnarray}

\noindent where $c=238$ m/s is calculated from the dispersion curve measured in neutron scattering experiments \cite{pitaevs} and $m$ is the mass of helium atom.

We note that the kinetic energy of the transit atom is comparable, by order of magnitude, to the oscillatory ground-state energy which this atom had inside the cage before the jump (the oscillatory
atom is in the ground state around $\sim 1$ K because $\hbar\omega$ is about 54 K in helium \cite{annett}). Using $v=a\omega$ (here and below we drop the lower index in $\omega_{\rm D}$ for brevity)
for the transit as before and the uncertainty relation for the transit atom localised within the distance $a$ during the transit process as

\begin{equation}
mva=ma^2\omega\sim\hbar,
\label{uncert}
\end{equation}

\noindent we see that the oscillatory energy in the ground state $E_g=\frac{\hbar\omega}{2}$ and the transit kinetic energy $E_t=\frac{mv^2}{2}$ become the same. This is consistent with the
oscillatory-transit dynamics envisaged by Frenkel \cite{frenkel}: an oscillating atom becomes a transit when the cage atoms move out of the way. In this process, the energy of the transit atom is
close to the kinetic energy this atom had inside the cage before the jump. This can be seen by recalling that the average kinetic energy of the oscillating atom inside the cage in the ground state,
$\left<K_g\right>$, is one half of the ground state energy: $\left<K_g\right>=\frac{\hbar\omega}{4}$. Taking $\hbar\omega$=54 K \cite{annett}, we find $\left<K_g\right>$ close to $E_t$ in
\eqref{moment}.

More generally, the closeness of the energy of flow-enabling liquid transits and the in-cage oscillatory energy is consistent with the increasingly appreciated similarity of liquid and solid properties and the general approach to liquids based on solidlike, rather than gaslike, concepts \cite{wallacebook,dyre,ropp,proctor2}. This similarity has remained under-explored in liquids \cite{ropp}. This notably included quantum liquids where gaslike approaches were used such as the Bose-Einstein condensation at zero momentum \cite{pinesnoz,griffin,pitaevs,annett,leggett}.

The above closeness of the transit energy and the energy this atom had inside the cage can be used to approximately estimate the distribution of transit momenta. This closeness implies that momenta of transit atoms are close to momenta these atoms had inside the cage, $p$ (except for small $p$ as discussed below) and, consequently, that the distributions of momenta of both sets of atoms are also close. The distribution of $p$ inside the cage is set by the momentum distribution of the harmonic oscillator in the quantum regime $T\ll\hbar\omega$ in the ground state, with the probability

\begin{equation}
\rho\propto e^{-\frac{p^2}{m\hbar\omega}}
\label{prob}
\end{equation}

Large $p$ give small $\rho$. Small $p$ do not apply to transits because flow-enabling transits are highly mobile by definition. Indeed, zero momenta of transits implies the absence of liquid flow. Lets consider $\rho$ at intermediate $p$ corresponding to $p=p_t=mc$ in \eqref{moment}. Using $c=a\omega$ for the transit atom as before, the argument in the exponent becomes $\frac{ma^2\omega}{\hbar}$ in absolute value. This ratio is on the order of 1 in view of the uncertainty relation \eqref{uncert}. Using $\frac{p_t^2}{2m}=14$ K and $\hbar\omega$=54 K as before, the argument in the exponent is about $-0.5$. Therefore, $\rho$ is not small, corresponding to a large number of transit atoms with momenta close to $p_t$. This is the case because the probability \eqref{prob} is quantum. Were the probability classical and set by the thermal distribution of high-energy transits with $p=p_t$ as

\begin{equation}
\rho\propto e^{-\frac{p_t^2}{2mT}},
\end{equation}

\noindent $\rho$ would be close to 0 because $\frac{p_t^2}{2m}\gg T$ at $\sim 1$ K.

We have seen that the two parameters characterising the condensed state of matter, $a$ and $\tau_{\rm D}$, constrain the speed of a large number of transits to be close to $v=\frac{a}{\tau_{\rm D}}=c$. Characteristic values of these two parameters, $a$ and $\tau_{\rm D}$, are set by fundamental physical constants \cite{sciadv1}. As a result, $c$ is also governed by fundamental physical constants \cite{sciadv2}. This additionally points to the physical significance of the state \eqref{moment}.

We now add the main proposal: BEC in liquid $^4$He corresponds to a large number of transits (e.g. tunneling transits) occupying the state with a finite momentum close to $p_t$ and energy $E_t$ in
Eq. \eqref{moment}. The BEC at finite energy (BECFE) takes place on a sphere with the radius close to $p_t=mc$ in momentum space (the corresponding wavevector $k_t$ is $k_t=\frac{mc}{\hbar}\approx
1.5$ ~\AA$^{-1}$) and the width set by fluctuations of transit speeds. The net momentum of all transits is zero because transits move in different directions (see Figure 1c). Similarly to the Fermi
sphere where $k_F\propto\frac{1}{a}$, $k_t=\frac{mc}{\hbar}=\frac{ma\omega}{\hbar}$, or

\begin{equation}
k_t\approx\frac{1}{a}
\end{equation}

\noindent where we used the uncertainty relation \eqref{uncert}.

The accumulation of transit boson atoms in the state close to \eqref{moment} can therefore be thought of as the emergence of a {\it Bose-liquid sphere} with radius related to interatomic spacing. This sphere exists in Bose liquids with high-energy mobile transits but not in Bose gases where atoms condense at the centre of the sphere with $p=0$. The transit states are within the sphere width rather than in the ball inside the sphere as is the case for Fermi particles.

The de Broglie wavelength of transits, $\lambda_{\rm dB}=\frac{h}{p_t}$, is above 4 \AA, implying that the transit wavefunctions overlap, sustaining a macroscopic wavefunction $\Psi_T$ which can be Bose-symmetrised. 


The BEC at a finite momentum was discussed in general terms \cite{pinesnoz}, developed for the case of finite momentum and zero energy \cite{yukalov} and applied to bosons in optical lattices and
thin films \cite{finite1,finite2}.

We note that this picture does not depend on whether the transit wavefunction $\psi$ is approximated by the plane wave or not. The predictions discussed below concern the transit energy $E_t\propto
p_t^2$ in Eq. \eqref{moment} which depends on $p_t=|{\bf p}_t|$ only and is the same if ${\bf p}_t$ changes direction during the transit process. We also note that it is the translational motion of
transits that enables their accumulation in the state with a certain momentum; the motion of the rest of the atoms is oscillatory inside the cages where momentum is oscillating and is not constant
during any time duration.

As discussed earlier, an oscillating atom in the ground state becomes the transit when the cage opens up due to fluctuations (e.g. when the atom tunnels, aided by cage fluctuations). The transit
energy is close to the kinetic energy of the zero-point motion this atom had inside the cage. The transit then settles into another cage where it again undergoes temporary oscillations, and so on.
Hence the transit energy is {\it periodically re-set} to the kinetic energy of the in-cage zero-point oscillatory motion $\frac{\hbar\omega}{4}=14$ K. This energy is well-defined when the in-cage
oscillatory component of motion exists. Therefore, re-setting of the transit kinetic energy, together with the finiteness and the constancy of the in-cage zero-point energy (and hence the existence
of cages where the zero-point oscillations take place), importantly contribute to the condensation of transits at a finite energy. In this sense, the zero-point energy in liquid helium has importance
beyond its usually discussed role \cite{annett,andreev}. We will return to this point in the Discussion section where we will discuss the transition to the state where cages disappear at high
temperature.

\subsection{Experimental data}
\label{exp}

\begin{figure}
\begin{center}
{\scalebox{0.35}{\includegraphics{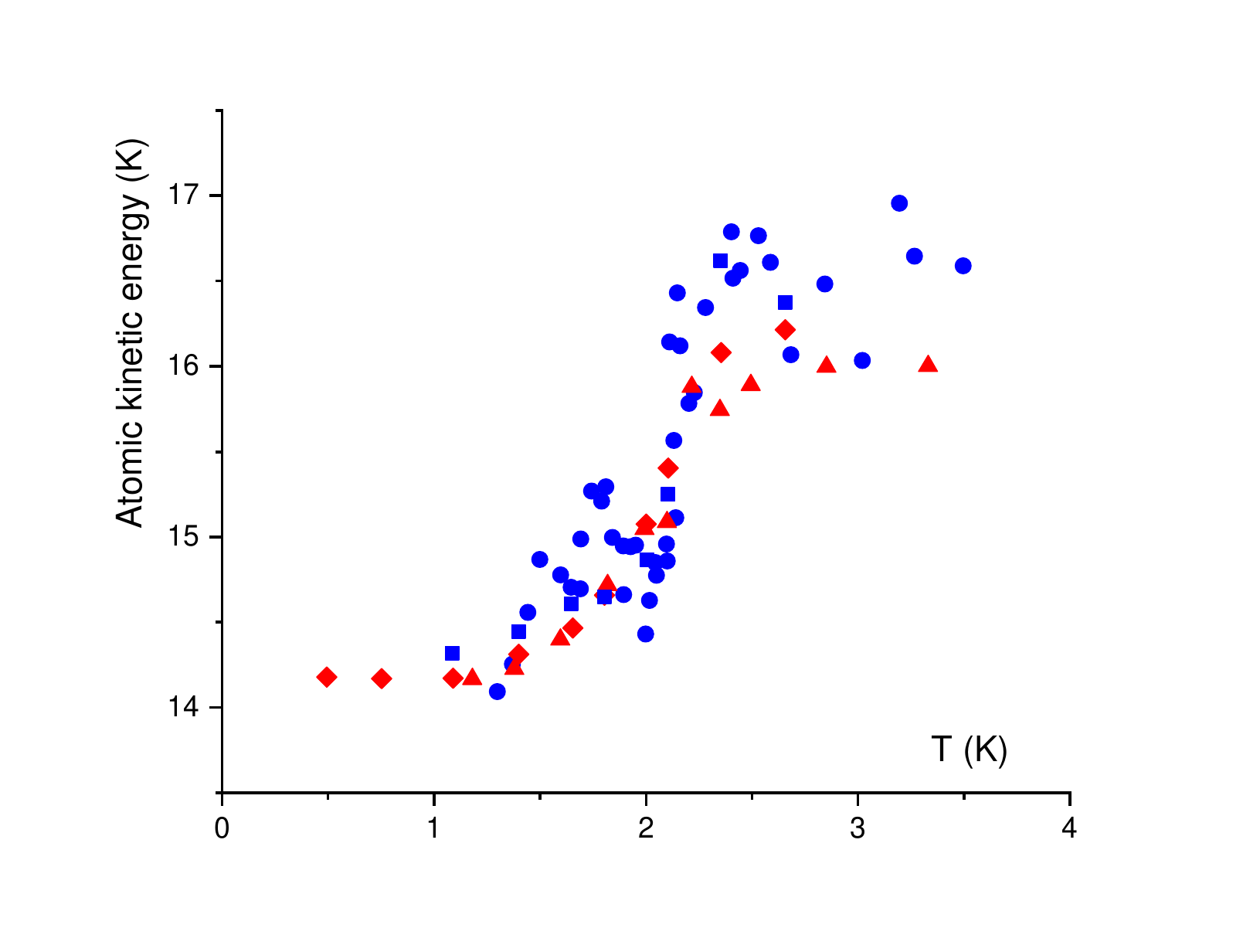}}}
\end{center}
\caption{Atomic kinetic energy measured in different sets of neutron scattering experiments in liquid $^4$He (circles) and calculated in path integral Monte Carlo simulations (triangles) at atmospheric pressure ($T_\lambda=2.17$ K) \cite{temp-energy3}. Squares correspond to more recent neutron experiments and diamonds to quantum Monte Carlo similations \cite{temp-energy4}. Experimental uncertainties of energy are about 0.2-0.6 K.}
\label{sqw1}
\end{figure}

The accumulation of a large number of transits in the state around $E=E_t$ should result in the average atomic kinetic energy $K$ to markedly change
below $T_\lambda$ and approach 14 K at low temperature as Eq. \eqref{moment} predicts. This is in agreement with neutron scattering experiments: as
shown in Figure \ref{sqw1}, this energy sharply reduces at $T_\lambda$ and approaches 14 K at low temperature
\cite{temp-energy2,temp-energy3,temp-energy4}.

We note that at low temperature, the experimental kinetic energy in Figure \ref{sqw1} is the average of the energy of the quasi-free translational motion of transits $E_t$ \eqref{moment} and the
average kinetic energy $\left<K_g\right>$ of atoms oscillating in cages in the ground state. These two energies are close to each other as discussed in the previous section, hence the experimental
kinetic energy of 14 K implies the average kinetic energy of transits to be also 14 K as Eq. \eqref{moment} predicts.

We also note that in the range $\Delta T=2.5$-3.5 K above $T_\lambda$, $K$ in Figure \ref{sqw1} is larger than $E_t=14$ K at low temperature by approximately the amount of the thermal energy
corresponding to $\Delta T$: $K=E_t+\Delta T$. Hence thermal energy is added to the particle zero-point energy (equal to the kinetic energy of transits). We will return to this point in the
Discussion section discussing transition to the state where cages disappear at high temperature.

Previously, the drop of $K$ at $T_\lambda$ was rationalised by assuming that a fraction of particles undergo BEC at $p=0$ and the rest having the energy in the uncondensed state \cite{temp-energy2,temp-energy3,temp-energy4}. Previous work did not explain the value of $K$ of 14 K below the jump at $T_\lambda$. This value readily comes out in the picture proposed here and from Eq. \eqref{moment}.


The experimental fluctuations of $K$ in Figure \ref{sqw1} are about $0.3$ K around 1 K \cite{temp-energy4}. Using $E_t$ in \eqref{moment}, this corresponds to fluctuations of the relative energy and momentum of about 1-2\%.

The second prediction concerns the dynamic structure factor $S({\bf q},\omega)$. The BECFE operates in a weakly-interacting gas of transits embedded in a quasi-rigid network of oscillating atoms (see Fig. 1c). The accumulation of transits in the state \eqref{moment} should result in an increasing sharp peak of $S({\bf q},\omega)$ at frequency $\omega_p$ corresponding to the excitation energy, as $S({\bf q},\omega)\propto\delta(\omega-\omega_p)$ \cite{pinesnoz}. Excitations in a weakly-interacting Bose system are given by quasiparticles with the Hamiltonian

\begin{eqnarray}
\begin{split}
& H=E_0+\sum_{p\ne 0}\epsilon_p b_p^+b_p \\
& \epsilon_p=\sqrt{u_s^2p^2+\left(\frac{p^2}{2m}\right)^2}
\end{split}
\label{excit}
\end{eqnarray}

\noindent where $b_p^+$ and $b_p$ are creation and annihilation operators of elementary excitations with energy $\epsilon_p$, $u_s$ is the speed of sound in the system and where $\epsilon_p$ correspond to single-particle excitations $\epsilon_p=\frac{p^2}{2m}$ at large $p$ \cite{landaustat,pinesnoz}. The accumulation of transits in the state $E=E_t$ should then result in a sharp peak of $S({\bf q},\omega)$ below $T_\lambda$ at $\nu\approx0.3$ THz as Eq. \eqref{moment} predicts.

\begin{figure}
\begin{center}
{\scalebox{0.35}{\includegraphics{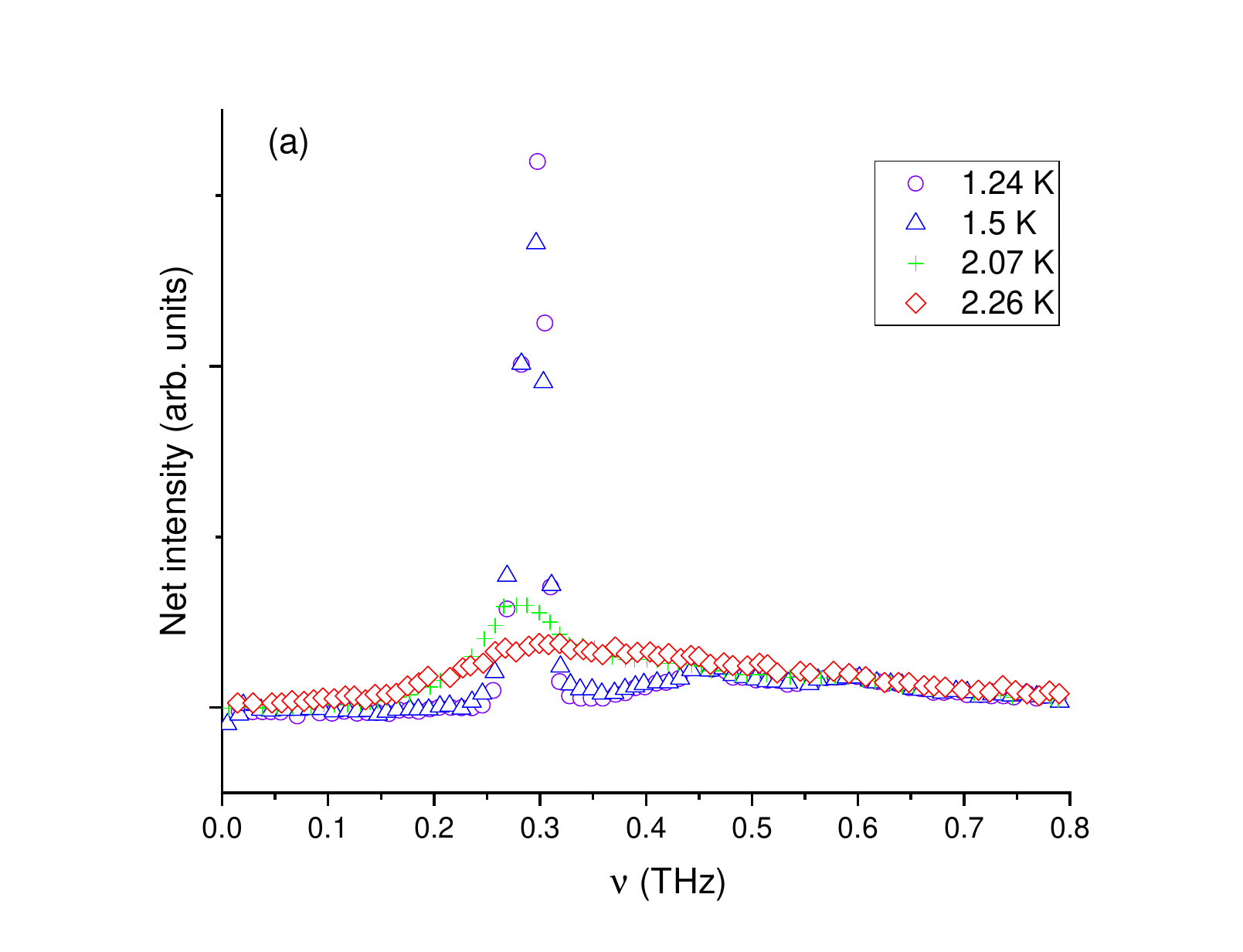}}}
{\scalebox{0.35}{\includegraphics{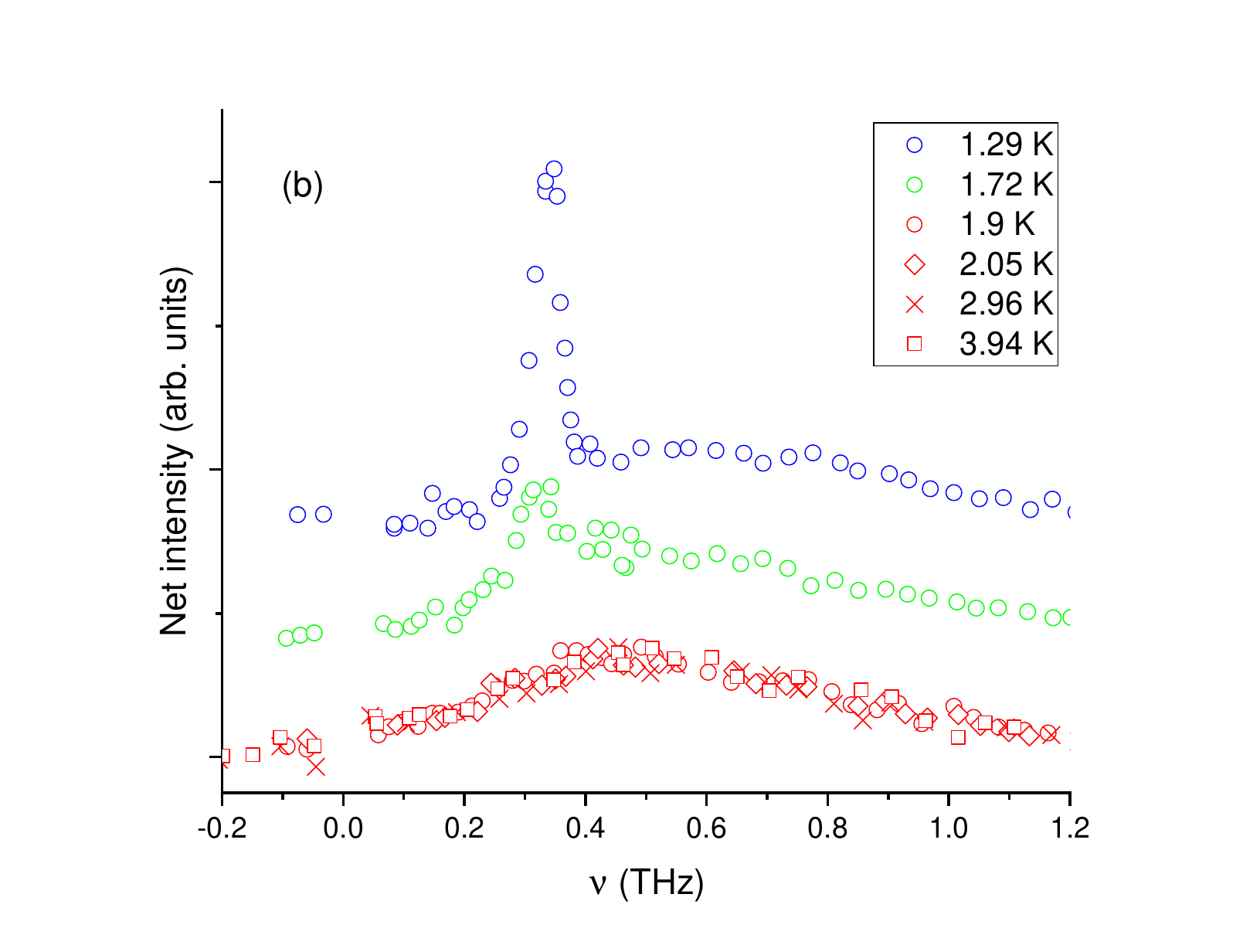}}}
\end{center}
\caption{(a) $S(q,\omega)$ measured in neutron scattering experiments at atmospheric pressure at $q=1$~\AA$^{-1}$. (b) $S(q,\omega)$ at 20 atm ($T_\lambda=1.928$ K) at $q=1.13$~\AA$^{-1}$. The data are from Refs. \cite{griffin,heliumpressure}.}
\label{sqw2}
\end{figure}

This prediction is supported by neutron scattering experiments
\cite{griffin,glyde,heliumneutron1,heliumpressure}. Figures \ref{sqw2}a,b
show broad frequency distribution above $T_\lambda$ and a remarkable
increase and sharpening of the peak below $T_\lambda$ (in some neutron
scattering experiments, the width of the low-temperature peak can not be
resolved \cite{glyde}). This is reminiscent of the sharp peak appearing in
gases undergoing BEC at $p=0$. The peak is close to 0.3 THz as Eq.
\eqref{moment} predicts.

We show the scattered intensity at $q\approx 1$~\AA$^{-1}$ because this $q$ and its vicinity correspond to single-particle excitations as follows from a detailed analysis of liquid helium spectra
(see, e.g., Ref. \cite{griffin}, pp. 20,167,181,193 etc). Sharp peaks at these $q$ are therefore consistent with our picture of transits in the BECFE state. Experimentally, sharp peaks are seen to
develop for large $q$ in this range as seen in Figure \ref{sqw2} but not at smaller $q$. For example, no sharp change is seen in the scattering intensity below and above $T_\lambda$ at
$q=0.4~$\AA$^{-1}-0.6~$\AA$^{-1}$ attributed to the phonon part of the spectrum (see, e.g., Ref. \cite{griffin}, pp. 157-159 etc). We note that at $q$ significantly exceeding those shown in Figure
\ref{sqw2}, the single-particle resonance no longer appears as a distinct peak due to experimental issues and other factors \cite{griffin}.


Previously, the sharp resonance-type peaks in Fig. \ref{sqw2}a,b were related to the BEC at $p=0$ which hybridizes single-particle excitations with collective density fluctuations \cite{griffin,glyde}. It was later noted that the evidence for the BEC at $p=0$ is inconclusive because (a) the experimental data analysis giving the number of atoms in the state $p=0$, $n_0$, requires the prior knowledge of a physical model and (b) the data can be fit by models where $n_0=0$ (see, e.g., p. 175 of Ref. \cite{temp-energy4} and accompanying discussion and references). Previous interpretations involved parameters fitted to experimental data and did not physically explain why the resonance takes place at about 0.3 THz. This value readily comes out in the picture proposed here and from Eq. \eqref{moment} in particular.


\subsection{Discussion}
\label{psi}

The transit subsystem is related to the liquid flow as discussed earlier. Its wavefunction $\Psi_T$ can be written by approximating the transit motion by plane waves as mentioned earlier. In a simple
model example where the transit wave function is oscillatory, its speed is given by $c$ and where the fluctuations of transit speeds are not considered, the space variation of $\Psi_T$ is

\begin{equation}
\Psi_T=\prod\limits_{i=1}^{N_t}\exp\left(i\frac{mc}{\hbar}r_i\right)
\label{mbody}
\end{equation}

\noindent where the single-particle functions are assumed normalised, $N_t$ is the number of transits in the BECFE state $p=p_t=mc$, $c=|{\bf c}_i|$ in Eq. \eqref{moment}, ${\bf c}_i$ is the transit velocity, $r_i=|{\bf r}_i(x,y,z)|$ ($0<r_i<a$) and ${\bf r}_i$ is the vector along the translational transit motion beginning at the transit starting point and defining the field of transit displacements in Figure 1c.

Although the product of plane waves in $\Psi_T$ may resemble the ideal gas, $\Psi_T$ describes a very different system: the gas of transits embedded in a quasi-rigid network of oscillating atoms in the liquid (this network is quasi-equilibrium and changes with relaxation time $\tau$ as discussed earlier). For this reason, the BECFE state does not suffer from the instability of the superfluid phase in the Bose gas related to the parabolic dispersion law and zero compressibility \cite{pinesnoz}: the overall system is the liquid with a finite compressibility and speed of sound $c$ where the BECFE operates in a subsystem of transits.

A question which perhaps needs further study is the dynamical nature of BECFE. The probability of an atom to be in the transit state is $\frac{\tau_{\rm D}}{\tau}$, where $\tau_{\rm D}$ is the time
this atom spends in the transit state and $\tau$ is the time between transits. In statistical equilibrium, the number of transits at each moment of time is

\begin{equation}
N_t=N\frac{\tau_{\rm D}}{\tau}
\end{equation}

\noindent where $N$ is the total number of atoms. 

Hence the BECFE corresponds to a fixed set of $N_t$ transits during time $\tau_{\rm D}$, as illustrated in Figure 1c. In the next time period $\tau_{\rm D}$, a different set of transits is formed.
After time $\tau$, all atoms participate in the transit motion and the BECFE. The liquid behavior and the superfluid response therefore correspond to a continuous time sequence of different transit
subsets. The BECFE operates in each subset and is described by $\Psi_T$ in \eqref{mbody}. This dynamical picture then addresses the long-standing problem of identifying a particular set of helium
atoms with the superfluid and with the BEC state \cite{landaustat,landaumigdal,pinesnoz,pitaevs}: all atoms participate in the superfluid response and BECFE state but they do so at different times.

A macroscopically-occupied condensate state is related to superfluidity since modifying this state involves a simultaneous action on a large number of transits in the symmetrised macroscopic wavefunction \cite{pinesnoz}. The spatially-varying condensate velocity, or the superfluid velocity, is ${\bf v}_s=\frac{\hbar}{m}\nabla S$, where $S$ is the phase of the condensate wavefunction $\psi_c$ \cite{pinesnoz}. From Eq. \eqref{mbody}, $\psi_c=\exp\left(i\frac{mc}{\hbar}r\right)$. This gives ${\bf v}_s=c\frac{{\bf r}}{r}={\bf c}_i$, whereas the net ${\bf v}_s$ is 0 because transits move in different directions. If the liquid moves with velocity ${\bf u}$, $\Psi_T$ in Eq. \eqref{mbody} becomes

\begin{equation}
\Psi_T=\prod\limits_{i=1}^{N_t}\exp\left(i\frac{m}{\hbar}({\bf c_i}+{\bf u}){\bf r}_i\right)
\label{mbody1}
\end{equation}

This gives ${\bf v}_s={\bf c}_i+{\bf u}$ and the net velocity of dissipationless flow of each transit subset ${\bf v}_s={\bf u}$. Experimentally, $v_s$ is limited by vortices \cite{pinesnoz,pitaevs}. If ${\bf u}=-{\bf c_i}$ and an atom is stationary, the single-transit wavefunction becomes a constant and is removed from the BECFE. This makes an interesting connection to the macroscopic Landau criterion for the critical velocity set by $c$ \cite{landaustat,pinesnoz}.



We make three remarks related to how the proposed picture is related to previous work. First, the transits do not contribute to the liquid specific heat or other derivatives of thermodynamic
potentials because $\exp(-E_t/T)$ in the partition function and its derivatives are negligibly small around 1 K. 
Recall that transits are athermal and that the transit energy comes from the zero-point oscillatory motion in a quasi-equilibrium cage: as the cage opens up, the oscillating atom becomes the transit
undergoing a translational (e.g. tunneling) motion between two neighbouring positions (see discussion after Eq. \eqref{moment}). Low-temperature thermodynamic properties of liquid helium are instead
related to low-frequency phonons. This gives the specific heat $c_v\propto T^3$ \cite{landaumigdal}, in agreement with experiments \cite{greywall}. The anomaly of $c_v$ at $T_\lambda$ was interpreted
as being due to the increasing size of permutation polygons corresponding to atomic exchanges \cite{feynman} (this picture does not consider BEC at $p=0$).

Second, the proposed picture addresses the issue related to BEC in the $p=0$ state in liquid helium raised early by Landau: the exchange of momentum between the immobile atoms in the state $p=0$ and
excited mobile ones would result in friction, precluding superfluidity \cite{landaumigdal}. This is not an issue in our picture where transits accumulate in their natural high-energy mobile states.

Third, earlier discussion of BEC in liquid helium at $p=0$ was based on the analogy with gases \cite{pinesnoz,griffin,ceperley,pitaevs,annett,leggett,glyde}. The data involved in this discussion are
not inconsistent with the BECFE considered here. Theoretically, BEC in liquid helium at $p=0$ was discussed by generalising the criterion for the Bose gas and based on a large density matrix
eigenvalue related to a large number of atoms in a particular state \cite{penrose}. This applies to BEC at any $p$ \cite{pinesnoz}. Experimentally, ascertaining BEC at $p=0$ involves decomposing the
measured momentum distribution into $n_0\delta(p)$ ($n_0$ is the number of atoms with $p=0$) and the rest, i.e. by assuming BEC at $p=0$ to begin with \cite{temp-energy4,griffin,pitaevs}. The
analysis of experimental data is model-dependent and involves problems in inverting scattering data to obtain a non-ambiguous momentum distribution. As a result, several models are consistent with
$n_0=0$ (see, e.g., p. 175 of Ref. \cite{temp-energy4} and accompanying discussion). By similarly assuming BEC at $p=0$, the condensate fraction was calculated in PIMC simulations \cite{ceperley}. It
would be interesting to use these as well as dynamical PIMD simulations \cite{path-he} to quantify the BECFE state at $E=E_t$ related to transits. This would require larger systems with many
transits.

We have not discussed the detailed nature of the transition at $T_\lambda$ itself, leaving it for future work. A useful insight comes from the dynamical crossover at the Frenkel line (FL) marking the
transition of particle motion from combined oscillatory motion and translational (diffusive) transit motion at low temperature below the line to purely transit motion above the line
\cite{flreview,Brazhkin2013}. As discussed in section \ref{dynamics}, the transit energy is {\it periodically re-set} to the kinetic energy of the in-cage zero-point oscillatory motion
$\frac{\hbar\omega}{4}=14$ K, and this re-setting importantly contributes to BECFE of transits. The zero-point oscillation energy is due to the presence of the in-cage oscillatory component and,
therefore, to the existence of cages themselves during at least some time longer than the shortest elementary vibration period $\tau_{\rm D}$. This corresponds to the dynamics below the FL
\cite{flreview,Brazhkin2013}. Above the FL, the in-cage oscillatory component of atomic motion disappears, and the cages themselves no longer exist even during $\tau_{\rm D}$
\cite{flreview,Brazhkin2013}. The dynamics of all atoms becomes purely translational. In this state, dynamics of transits is largely affected by their mutual interactions and random local
configurations (but not by the above constraining cage effect below the FL), to which disordering temperature fluctuations start contributing (see Fig. 2 and its discussion in Section \ref{exp}). As
a result, the distribution of atom momenta is expected to widen. Hence the transition to the BECFE state and the superfluid state can be provisionally related to the transition at the FL.

The temperature at the FL, $T_F$, corresponds to the disappearance of the minimum of the velocity autocorrelation function (VAF) \cite{Brazhkin2013} (here, we consider the quantum analogue of the
FL). Quantum path integral simulations in liquid helium accounting for the particle exchange effects show that the minimum of VAF disappears between 2.5 K and 1.18 K (see Fig. 3 of Ref.
\cite{path-he}). Hence the temperature at which the minimum disappears can be estimated as the average of these two temperatures (in the future it would be interesting to calculate the VAF at a finer
temperature grid to establish $T_F$ more precisely). This gives

\begin{equation}
T_F=T_\lambda=1.84~{\rm K}
\label{tf}
\end{equation}

This is in 15\% agreement with the experimental $T_\lambda=2.17$~K. This is a fairly good evaluation of $T_\lambda$ on theoretical grounds.

The evaluation of $T_\lambda$ is useful in view of the absence of a microscopic theory of liquid helium \cite{pinesnoz}, explaining why no such evaluation was made despite large and long-lasting body
of research in the area \cite{landaustat,pitaevs,annett,leggett,griffin}.

We note that $T_F$ and $T_\lambda$ are probably higher than the average of 2.5 K and 1.18 K in Eq. \eqref{tf} because the VAF at 2.5 K is close to developing a minimum, whereas the minimum of VAF at
1.18 K is fairly deep \cite{path-he}. This reduces the difference between predicted and experimental $T_\lambda$ to less than 15\%.

\section{Summary}

In summary, we discussed details of microscopic dynamics in liquids and explored its implications in liquid helium. This gives a way to discuss BEC in liquid helium which is consistent with
microscopic liquid dynamics. In this picture, high mobility of liquid atoms results in the accumulation of transits in a finite-energy state. This is consistent with the experimentally measured
kinetic energy of helium atoms below $T_\lambda$ and sharp peaks of scattered intensity at energy predicted by Eq. \eqref{moment}. The superfluid transition temperature is evaluated to be within 15\%
of the experimental value. More work is clearly needed to discuss the wealth of effects in liquid helium in this picture, including details of transition at $T_\lambda$.

\section{Acknowledgements}

I am grateful to A. E. Phillips and J. C. Phillips for discussions and EPSRC for support.


\section*{Appendix}

Below is an example of a quantum-mechanical model of oscillatory-transit dynamics with continuous solutions in space and time. Other examples can be found too involving model modifications.

Let us consider how the wave function of the atom undergoing the transit changes from one oscillatory state to the next as shown in Figure 1a. We choose the $x$-direction along the transit motion. The wave function $f$ of the oscillating atom around $\sim 1$ K can be written as the oscillator wavefunction in the ground state because $\hbar\omega$, where $\omega$ is oscillation frequency, is about 54 K as calculated from the interatomic helium potential \cite{annett}: $f\propto \exp(-\alpha x^2)$, where $\alpha=\frac{m\omega}{2\hbar}$ and $m$ is the mass of helium atom. The wavefunction of the transit $\psi(x)$ can be approximated by the plane wave as discussed in the main text. If $f_1(x)$ and $f_2(x)$ are the ground-state oscillatory wavefunctions before and after the transit, we can try the following wavefunctions describing the atom before, during and after the transit:

\begin{eqnarray}
\begin{split}
& f_1(x)=A\exp(-\alpha x^2)\\
& \psi(x)=B\sin(kx-\phi)\\
& f_2(x)=C\exp(-\alpha (x-a)^2)
\end{split}
\label{psi1}
\end{eqnarray}

Let $\delta$ be the distance at which the particle leaves the first oscillatory state and becomes the transit as shown in Figure 1b and $a-\delta$ the distance at which the transit settles in the second oscillatory state, where $a$ is the interatomic separation and $\delta<\frac{a}{2}$. The continuity of the wavefunctions in \eqref{psi1} is ensured by equating $f_1(x)$ and $\psi(x)$ and their derivatives at $x=\delta$ and equating $\psi(x)$ and $f_2(x)$ and their derivatives at $x=a-\delta$. This gives

\begin{eqnarray}
\begin{split}
& A\exp(-\alpha\delta^2)=B\sin(k\delta-\phi)\\
& -2A\alpha\delta\exp(-\alpha\delta^2)=Bk\cos(k\delta-\phi)\\
& C\exp(-\alpha\delta^2)=B\sin(k(a-\delta))-\phi)\\
& 2C\alpha\delta\exp(-\alpha\delta^2)=Bk\cos(k(a-\delta)-\phi)
\end{split}
\label{psi2}
\end{eqnarray}

The first and second pair of equations have non-trivial solutions each if

\begin{eqnarray}
\begin{split}
& -2\alpha\delta=\frac{k\cos(k\delta-\phi)}{\sin(k\delta-\phi)} \\
& 2\alpha\delta=\frac{k\cos(k(a-\delta)-\phi)}{\sin(k(a-\delta)-\phi)}
\end{split}
\label{psi3}
\end{eqnarray}


The two equations \eqref{psi3} are compatible if $\tan(k\delta-\phi)=-\tan(k(a-\delta)-\phi)$, giving, bar the periodic argument $\pm\frac{\pi n}{2}$, $\phi=\frac{ka}{2}$. Using this $\phi$ in \eqref{psi3} gives

\begin{eqnarray}
\begin{split}
& \tan y=\chi y\\
& y=k\left(\frac{a}{2}-\delta\right)\\
& \chi=\frac{1}{2\alpha\delta\left(\frac{a}{2}-\delta\right)}=\frac{a^2}{\delta\left(\frac{a}{2}-\delta\right)}
\end{split}
\label{tany}
\end{eqnarray}

In the last equality, we recalled $\alpha=\frac{m\omega}{\hbar}$ and used the uncertainty relation for the transit localised within distance $a$ in Eq. \eqref{uncert}, resulting in $\alpha=\frac{1}{a^2}$.

The equation for $y$ has a solution in the first half-period of $\tan y$ because $\chi>1$ in the considered range $0<\delta<\frac{a}{2}$.



In addition to the continuity of the wavefunctions in space, this model also ensures their time continuity. The time dependence of the three wavefunctions is

\begin{eqnarray}
\begin{split}
& f_1(x,t)\propto f_1(x)\exp\left(-\frac{iE_gt}{\hbar}\right)\\
& \psi(x)\propto\phi_1(x)\exp\left(-\frac{iE_tt}{\hbar}\right)\\
& f_2(x,t)\propto f_2(x)\exp\left(-\frac{iE_gt}{\hbar}\right)
\end{split}
\label{psi4}
\end{eqnarray}

\noindent where $E_g=\frac{\hbar\omega}{2}$ is the oscillatory ground state energy and $E_t=\frac{p^2}{2m}$ is the transit kinetic energy.

Setting $v=a\omega$ as earlier, $E_g$ and $E_t$ become equal: $E_t=\frac{m\omega^2a^2}{2}$ becomes $E_g=\frac{\hbar\omega}{2}$ once the uncertainty relation \eqref{uncert} is used, implying the time continuity of the wavefunctions in the oscillatory-transit states.




\end{document}